\documentclass{aa}  
\usepackage{graphicx}
\usepackage{longtable,lscape}
\usepackage[authoryear]{natbib}
\bibpunct{(}{)}{;}{a}{}{,}
\usepackage{txfonts}
\begin{document}
   \title{New very massive stars in Cygnus OB2} 
\author{I.~Negueruela
          \inst{1,2}
          \and
A.~Marco\inst{1,2}
\and
A.~Herrero\inst{3,4}
\and
J.~S.~Clark\inst{2}
  % \offprints{I.~Negueruela}
   \institute{Departamento de F\'{\i}sica, Ingenier\'{\i}a de Sistemas y
  Teor\'{\i}a de la Se\~{n}al, Universidad de Alicante, Apdo. 99,
  03080 Alicante, Spain\\
              \email{ignacio@dfists.ua.es}
         \and
	  Department of Physics and Astronomy, The Open University,
  Walton Hall, Milton Keynes MK7 6AA, United Kingdom 
\and
Instituto de Astrof\'{\i}sica de Canarias, 38200 La Laguna, Tenerife, Spain
\and
 Departamento de Astrof\'{\i}sica, Universidad de La Laguna, Avda.
Astrof\'{\i}sico Francisco S\'anchez, s/n, E-38071 La Laguna, Spain
}} 
   \date{Received }

% \abstract{}{}{}{}{} 
% 5 {} token are mandatory
 
  \abstract{The compact association Cygnus OB2 is known to contain a large
  population of massive stars, but its total mass is currently a matter of debate.
 While recent surveys have uncovered large numbers of OB stars
  in the area around Cyg~OB2, detailed study of the optically brightest
  among them suggests that most are not part of the association.}
   {We observed an additional sample of optically faint OB star candidates,  with the aim of checking
  if more obscured candidates are correspondingly more likely to be members of
  Cyg~OB2.} 
   {Low resolution spectra of 9 objects allow the rejection of one
  foreground star and the selection of four O-type stars, which were
  later observed at higher resolution. In a subsequent run, we
  observed three more stars in the classification region and three
  other stars in the far red.}
   {We identify five (perhaps six) new evolved very massive
     stars and three 
     main sequence O-type stars, all of which are likely to be members
     of Cyg~OB2. The new findings allow a much better definition of
     the upper HR diagram, suggesting an age $\sim2.5$~Myr for the
     association and hinting that the O3--5 supergiants in the
     association are blue stragglers, either younger or following a
     different evolutionary path from other cluster members. Though the bulk of the early stars seems to
     belong to an (approximately) single-age population, there is
     ample evidence for the presence of somewhat older stars at the
     same distance.}  
   {Our results suggest that, even though Cyg~OB2 is unlikely to
  contain as many as 100 O-type stars, it is indeed substantially more
  massive than  was thought prior to  recent infrared surveys.}


   \maketitle
%
%________________________________________________________________

\section{Introduction}
Among Galactic OB associations, Cyg OB2 is special in many
respects. For a start,  it is known to host a large population of
massive stars, including a significant fraction of the earliest
spectral types in the Galaxy \citep{wal02}. The optical extinction to Cyg OB2
is high, but not sufficiently so that it prevents spectra of its
stars in the classification region being taken (something impossible
for  other very massive open clusters with a large population of
massive stars, such 
as Westerlund~1 \citep{clark05} or the Arches Cluster
\citep{fig02}). Because of this, Cyg~OB2 is a very useful laboratory,
since, on  one hand, it provides a large homogeneous population
of OB stars that can be analysed \citep{her99,her02} and, on the other,
can be used as a template to compare optical and infrared
investigations \citep[e.g.,][]{hanson}. Finally, because of its
  compactness and high stellar content, Cyg~OB2 seems to 
occupy a more or less unique position somewhat 
intermediate between an open cluster and a normal OB association
\citep[cf.][]{kno00}. 
 
These properties have led to a great deal of interest in Cyg~OB2, from
the ``classical'' study of \citet{jm54} to the comprehensive
investigation by \citet{mt91}, who identified $\sim60$ stars more
massive than $15M_{\sun}$. More recently,  based on star
counts in the 2MASS observations of the region, \citet{kno00}  proposed that
the number of O-type stars in Cyg OB2 was much larger.
Building on this result, \citet{com02} preselected a large number of
possible OB members of Cyg~OB2 from their 2MASS colours and obtained
low-resolution $H$- and $K$-band spectroscopy of the
candidates. Candidates that lacked molecular bands were selected as
very likely early-type stars. Of 77 candidates so selected, 31 stars
for which optical spectra existed were OB stars, suggesting that most,
if not all, of the other 46 objects were also OB stars in Cyg OB2.

From this list of candidates,
\citet{hanson} selected those brightest in the optical (14 objects with $B= 
12$ to 14), for which she obtained classification spectra, finding
that all of them were indeed  OB stars. However, \citet{hanson} argues
that most of these objects are not members of Cyg~OB2. For a start,
they all lie at some distance from the previously defined boundaries
of Cyg~OB2, as most of the sources located by \citet{com02}
do. Moreover, about half of the objects observed are late O and early
B supergiants, indicating ages rather larger than the 2~Myr that
\citet{hanson} derives for Cyg~OB2 from  isochrone fitting to the
location of the main sequence. Finally, one star (A39, B2\,V) appears
far too bright for its spectral type and is almost certainly a  foreground
object.

It is therefore an open question as to whether the list of candidates from
\citet{com02} really contains a high fraction of actual Cyg~OB2
members. Here we investigate this issue with new spectra of several
other fainter optical candidates. We also make use of
the recent publication of a large catalogue of accurate spectral types
for Cyg~OB2 members \citep{kiminki}, which combined with our results
and those of \citet{hanson}, allows an enormous improvement in the
characterisation of the HR diagram for the association.

In what follows, we will use the notation of \citet{com02} for stars
within their list (A\#\# for OB candidates and B\#\# for emission-line
stars). For other members, we will use the numbering system of
\citet{mt91}, with prefix MT, except for the twelve stars with the
classical numbering of \citet{jm54}, which are given with the
symbol \# followed by their number.

%__________________________________________________________________

\section{Observations}

Candidate stars from the list of \citet{com02} were observed with the 1.52-m
G.~D.~Cassini telescope at the Loiano Observatory (Italy) during the
nights of 2004 July 15 -- 18. The
telescope was equipped with the Bologna Faint Object Spectrograph and
Camera (BFOSC) and an EEV camera. We used grism \#3, which covers
3300--5800\AA\ with a resolution of $\sim6$\AA.
Unfortunately, on the night of July 16th, the sky was very poor, with
some veiling, and we resorted to observing two stars with the
lower-resolution grism \#4. The night of July 
17th we could not observe. Therefore, in total, we observed only 10
stars, of which one, A27, had been observed before with better
resolution and signal-to noise ratio (SNR) by \citet{hanson}.

From these ten objects we selected five to be observed at higher
resolution; the four which appeared to be O-type stars (based on the
analysis presented in Section~\ref{loiano}) and one that
was probably a B-type dwarf as a check. These objects were observed
with the 4.2-m William Herschel Telescope (WHT) in La
Palma (Spain), equipped with the ISIS double-beam
spectrograph, during a service run in June 2006. The instrument was
fitted with the R300R grating and 
MARCONI2 CCD in the red arm and the R300B grating and EEV\#12 CCD in
the blue arm. Both configurations result in a nominal dispersion of
0.85\AA/pixel (the resolution element is approximately 3 pixels in the
blue an 2 pixels in the red).

As the selection criteria of \citet{com02} proved sound, we then
selected some objects from their list with very bright $K$ magnitudes
(which should be intrinsically brightest) and observed them during a
run on 2007 August 21-22 at the WHT. Three objects were observed in
the blue with grating R1200B (nominal dispersion of
$\sim0.23$\AA/pixel) and three others (whose $B>16$ made too faint for
the blue grating) were observed with the red arm and grating R600R in
the $I$-band, where relatively accurate classification is also
possible \citep[e.g.,][]{clark05}. This configuration has a nominal
dispersion of $\sim0.5$\AA/pixel.
 
All the spectra have been reduced with the {\em Starlink}
packages {\sc ccdpack} \citep{draper} and {\sc figaro}
\citep{shortridge} and analysed using {\sc figaro} and {\sc dipso}
\citep{howarth}. 

\section{Results}

\subsection{Loiano spectra}
\label{loiano}

\begin{figure}
   \centering
   \resizebox{\columnwidth}{!}{\includegraphics{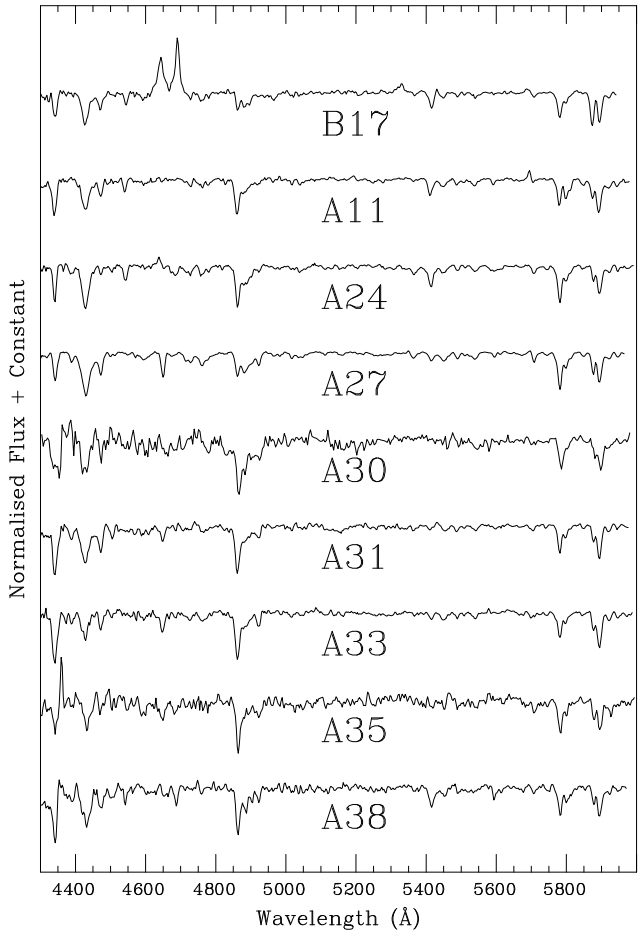}}
   \caption{Useful section of the spectra obtained from Loiano. The
     SNR decreases quickly towards the blue and there are essentially
     no counts bluewards of $\sim4200$\AA.The spectrum of A40, which
     turned out to be a 
   foreground star, is not shown.}
              \label{lowres}
    \end{figure}

The Loiano spectra have rather poor SNR in the blue, but allow a rough
classification of the stars. One of the candidates, A40, turns out to be a
foreground G-type star. The other 9 objects are very obviously OB
stars. Their spectra are displayed in Fig.~\ref{lowres}, while their
2MASS magnitudes and derived spectral types are listed in
Table~\ref{objects}. 

%---------------------------------------------------
   \begin{table}[ht]
\begin{center}
      \caption[]{Infrared 2MASS photometry and derived spectral types
        for programme stars.} 
         \label{objects}
\begin{tabular}{l c c c c}        % centered columns (5 columns)
\hline\hline \noalign{\smallskip}
   
Name & $(J-K_{{\rm S}})$ & $K_{{\rm S}}$& Spectral type & Telescope       \\
            \hline\noalign{\smallskip}

A11 & 1.19 & 6.64 &O7.5\,Ib-II(f)& L, WHT1\\
A12 & 1.21 & 5.72 &B0\,Ia & L,WHT2\\
A15 & 1.14 & 6.81&O7\,Ib(f)&L,WHT2\\
A18 & 1.07 & 8.35&$\sim$O8\,V& WHT3 \\
A24 & 0.97 & 7.46 &O6.5\,III((f))& L,WHT1\\
A25 & 1.01 & 7.36&$\sim$O8\,III & WHT3\\
A26 & 0.97 & 8.19&O9.5\,V & L, WHT2\\
A27 & 0.97 & 5.75&$\sim$B0\,I$^{a}$& L\\
A30 & 0.81 & 8.61&$\sim$B2\,V& L\\
A31 & 0.95 & 7.98&$\sim$B0.5\,V& L\\
A33 & 0.87 & 8.60&B0.2\,V& L, WHT1\\
A35 & 0.81 & 8.47&$\sim$B0\,V& L\\
A38 & 0.85 &8.56 &O8\,V& L, WHT1\\
\hline
B10 & 1.45 & 8.12 &Be& WHT3\\
B17 & 1.21 & 6.44&Ofpe&L,WHT1,WHT3\\

            \hline
        \end{tabular}
\begin{list}{}{}
\item[]$^{(a)}$ B0\,Ia \citep{hanson}
\item[]Key for telescope configurations:\\
L -- Loiano Cassini Telescope, WHT1 -- WHT in 2006 with blue arm, WHT2 --
WHT in 2007 with blue arm, WHT3 -- WHT in 2007 with red arm ($I$-band only)
\end{list}
\end{center}
   \end{table}
%---------------------------------------------------

A11 has \ion{He}{i}~4471\AA\ $\approx\:$\ion{He}{ii}~4542\AA, no visible
\ion{He}{ii}~4686\AA\ (at this resolution; we see it in the WHT
spectrum) and \ion{C}{iii}~5696\AA\ strongly in emission. It is 
thus an $\sim$O7 supergiant. A24 has  \ion{He}{i}~4471\AA\
$<\:$\ion{He}{ii}~4542\AA\ and \ion{He}{ii}~4686\AA\ in
absorption and so it is a relatively unevolved mid O-type star. A27 has
very prominent \ion{C}{iii}~4650\AA\, no 
\ion{He}{ii}~4686\AA, weak \ion{He}{ii}~4512\AA, strong \ion{He}{i}
lines and very weakened
H$\beta$. It should be a $\sim$B0 supergiant, and indeed it has been
classified as B0\,Ia by \citet{hanson}, based on higher quality
spectra. 

The spectrum of A30 has lower resolution and SNR than the rest, but
extends into the red. The lack of \ion{He}{ii}~4512\AA\ makes it later
than B0, while the fact that H$\alpha$ is much deeper than the
\ion{He}{i}~6678, 7065\AA\ lines suggests that it is a mid B star
(e.g., later than $\sim$B2V).

A31 has moderately strong \ion{C}{iii}~4650\AA\ and very weak
\ion{He}{ii}~4686\AA\ and \ion{He}{ii}~4512\AA, suggesting a main
sequence star in the B0-1 range. A33 is similar, with a slightly
stronger \ion{C}{iii}~4650\AA, perhaps suggesting a higher
luminosity. A35 has stronger \ion{He}{ii}~4686\AA, but is unlikely to
be much earlier, as \ion{He}{ii}~4512\AA\ is weak.

A38 has moderately strong \ion{He}{ii} lines, but
\ion{He}{i}~4471\AA\ $>\:$\ion{He}{ii}~4542\AA, suggesting a late
O-type star, while the lack of emission lines indicates a low
luminosity.  Finally B17 is characterised by strong emission lines of
\ion{He}{ii}~4686\AA\ and \ion{N}{iii}, and may be an extreme Of
supergiant or an Ofpe/WNL star.
  
%______________________________________________ 
%
%______________________________________________ 
%

\subsection{WHT spectra}
\label{lapalma}

\begin{figure}
   \centering
   \resizebox{\columnwidth}{!}{\includegraphics[angle =-90]{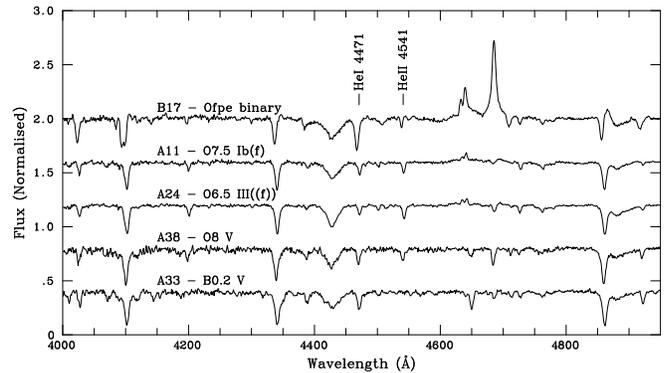}}
   \caption{Classification spectra of the 5 objects observed with ISIS
     on the
     WHT in June 2006.}
              \label{wht06}
    \end{figure}

Figure~\ref{wht06} shows the spectra of the 5 objects observed in
2006. The spectrum of B17 is very striking, with very strong
\ion{He}{ii}~4686\AA\ and \ion{N}{iii} emission, and a P-Cygni
profile in H$\beta$. All its lines are displaced by $>200\:{\rm
  km}\,{\rm s}^{-1}$ with respect to other members and show an
enormous shift in radial velocity with respect to the 2004
spectrum. We classify this 
object as an Ofpe star, almost certainly a binary, and will study it
in detail in a future paper. A11 and A24 have \ion{He}{ii}~4542\AA $\simeq$
\ion{He}{i}~4471\AA\ and are therefore close to O7, while the
in-filling of \ion{He}{ii}~4686\AA\ indicates a moderate
luminosity. A11 has \ion{N}{iii}~4630\,--\,4640\AA\ in emission, a
wind feature typical of luminous stars. Based on the criteria laid out
by \citet{wf90}, we 
classify A11 as O7.5\,Ib-II(f) and A24 as O6.5\,III((f)). 
A38 has \ion{He}{ii}~4686\AA\ strongly in absorption and we classify
it as O8\,V, though it is close to O8.5\,V, if we use the quantitative
criteria of \citet{mathys}. Finally, A33 has weak
\ion{He}{ii}~4686\AA\ and 4542\AA, but no \ion{He}{ii}~4200\AA, and
thus we classify it as B0.2\,V. The accurate classifications agree quite
well with the estimates obtained from the low-resolution spectra in
the previous section.

%______________________________________________ 
%
%______________________________________________ 
%

\begin{figure}
   \centering
   \resizebox{\columnwidth}{!}{\includegraphics[angle =-90]{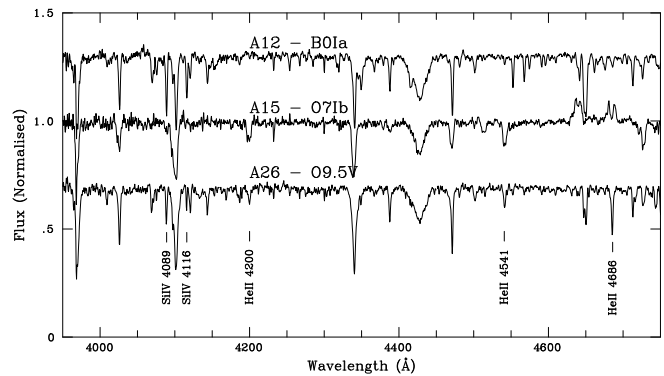}}
   \caption{Classification spectra of the 3 objects observed with the
     blue arm of ISIS on the
     WHT in August 2007. The most relevant lines are marked. }
              \label{wht07}
    \end{figure}

Figure~\ref{wht07} shows the 3 classification spectra obtained in
August 2007. The extremely prominent \ion{Si}{iv} lines in A12 show it to be
a luminous supergiant, while their ratio to \ion{Si}{iii} lines puts
it at B0, in agreement with the presence of three weak \ion{He}{ii}
lines. We adopt B0\,Ia. A15 is similar to A11 and
A24. \ion{He}{ii}~4686\AA\ is more clearly in emission, but the lack of
wind \ion{S}{iv} emission lines and weak
\ion{Si}{iv}~4089\AA\ prevent us from assigning a high luminosity. We
settle for O7\,Ibf. Finally, though clearly an O-type star because of
the strong \ion{He}{ii} lines, A26 still shows many weak \ion{Si}{iii}
and \ion{O}{ii} lines and is therefore O9.5\,V.

%______________________________________________ 
%
%______________________________________________ 
%

We also obtained red spectra of 4 objects, covering the atmospheric
window between 8300\AA\ and 8900\AA, where spectral classification is
possible \citep[e.g.,][]{clark05}. Fig.~\ref{fig:iband} shows the
spectra of three objects, B17 and two stars not observed in the blue,
A18 and A25. The three spectra are similar and typical of late O-type
stars. Only a few Paschen lines are visible, but the
\ion{C}{iii}~8502\AA\ line is clearly visible. This places the stars
in the O7--O9 range. The broadness of the Paschen lines and the small
number visible indicates that A18 and A25 are not supergiants. We take
an approximate spectral type $\sim$O8. Based on their $K$ magnitudes
and positions in the HR diagram (Section~\ref{sec:hr}), A25 is likely a
main-sequence object, but A18 could be a more evolved star.

A fourth object, B10 = MT 285, was also observed and its spectrum is
displayed in Fig.~\ref{fig:be}. The strong asymmetric emission Paschen
lines are typical of a Be star, with the very prominent
\ion{O}{i}~8446\AA\ indicates that it is not a late-B object
\citep{andrillat}. The likely detection of \ion{He}{i}~8779\AA\ and
strong \ion{O}{i}~7774\AA\ (not shown) emission indicates that it is
B2 or earlier \citep{andrillat}. This is fully consistent with the
detection by \citet{com02} of \ion{He}{i}~2.058$\mu$m in emission, as
this is only seen in Be stars earlier than B3 \citep{cs00}. Therefore
it is likely to be a massive Herbig Be 
star in Cyg~OB2. None of the classical Be stars observed by
  \citet{andrillat} shows $EW_{\ion{O}{i}}<-5.5$, but B10 has
  $EW_{\ion{O}{i}}\approx-10$ after correction for Pa~18, strongly
suggesting that it is a Herbig Be object, as they tend to have
stronger emission features.

%---------------------------------------------------
   \begin{table*}[ht]
\begin{center}
      \caption[]{Astrophysical parameters of programme stars, derived from model fits.}
         \label{params}
\begin{tabular}{l c c c c c c c c c c}        % centered columns (5 columns)
\hline\hline \noalign{\smallskip}
   
Name & Spectral type &$M_V$ &$T_{{\rm eff}}$&  $\log g$ &  $R $&
$\dot{M}$ & $v_{\infty}$ &$v_{{\rm rot}}$& Mass&$\log (L/L_{\sun})$\\
&& &(K)& & $(R_{\sun})$& ($M_{\sun}$)& (km~s$^{-1}$) & (km~s$^{-1}$)& ($M_{\sun}$)& \\
            \hline\noalign{\smallskip}

A11 & O7.5\,Ib-II(f)&$-5.8$&36000 &  3.6 &  15.9 & $2.2\times10^{-6}$& 1900 &$<160$& 38.9&5.6\\
A12 & B0\,Ia &$-6.7$&27000& 3.0&30.2&$3.5\times10^{-6}$&1350&80&34.2&5.6\\
A15 & O7\,Ibf&$-5.7$&35000&3.2&15.6&$3.2\times10^{-6}$&2100&245&19.0&5.5\\
A24 & O6.5\,III((f))&  $-5.0$&37500 & 3.6 &  10.7 & $1.7\times10^{-6}$&  2600&$<160$&18.1&5.3\\
A26 & O9.5\,V&$-4.2$&35000& 3.9&7.7&$4.1\times10^{-8}$&1300&90&17.6&4.9\\
A33 & B0.2\,V&$-3.6$ & 31000 &  4.0 &  6.6 &  $2.0\times10^{-8}$ &  1000&$<160$&16.6&4.6\\
A38 & O8\,V&$-3.7$&36000  &4.0   &6.0  &$4.9\times10^{-8}$ &1900&$<160$&13.8&4.7\\

            \hline
        \end{tabular}
\begin{list}{}{}
\item[]All models have been calculated using $\beta=0.8$, $v_{{\rm
    turb}}=10\:{\rm km}\,{\rm s}^{-1}$ and $\epsilon=0.09$, except for
  A12, which, being a B-type supergiant, required different wind
  parameters ($\beta=1.0$, $v_{{\rm turb}}=15\:{\rm km}\,{\rm 
  s}^{-1}$)  and A15, which required $\epsilon=0.25$.
\end{list}
\end{center}
   \end{table*}

%Finally we observed the well-known Blue HyperGiant (BHG) \#12. This
%object is a spectrum variable \citep{kiminki} and should perhaps be
%classified as 
%an LBV. In our spectrum, \ion{N}{i} lines are almost absent, with only
%a weak 8680--86\AA\ blend observed. This sets the spectral type at
%approximately B2.5. \ion{O}{i}~8446\AA\ has obvious P-Cygni emission
%(like the most 
%luminous B supergiants in Westerlund~1; \citealt{clark05}) and cannot
%be used to constrain the spectral type. In any case, the star should
%be classified as earlier than B3, but not earlier than B2, in clear
%contrast to classification at other epochs as B5\,Iae or even B8\,Iae
%\citep[cf.][]{kiminki}. 

%---------------------------------------------------

\subsection{Model fits}
The WHT spectra, even if of moderate resolution, offer a good chance
to complement the study presented by \citet{her99,her02}, expanding the
sample that can be analysed. For all stars with blue WHT
spectra except B17, which is unlikely to be a single star, we
determined stellar parameters using {\sc fastwind}
\citep{santo97,puls05}, by fitting H and He line profiles
in the standard way \citep{her92,rep04}.
The value of $v_{\infty}$ was adopted from the spectral type, after
\citet{kp00}. We adopted a value of the microturbulence $\xi=10\:{\rm
  km}\,{\rm s}^{-1}$ for all objects except A12
(the only B-supergiant in the sample) for which we adopted $15\:{\rm
  km}\,{\rm s}^{-1}$. $\beta$ (the
exponent of the velocity law) was adopted to be 0.8 and varied when the
fit could be improved. Again, only A12 needed a slightly larger value of $\beta$
(consistent with a slower wind acceleration), and we adopted
$\beta=1.0$ for this source
(but note that lacking H$\alpha$ or having low resolution, our data are not very
sensitive to $\beta$). We should indicate that the final fit to the
\ion{He}{ii}~4686\AA\ line of A12 is not satisfactory.

\begin{figure}
   \centering
   \resizebox{\columnwidth}{!}{\includegraphics[]{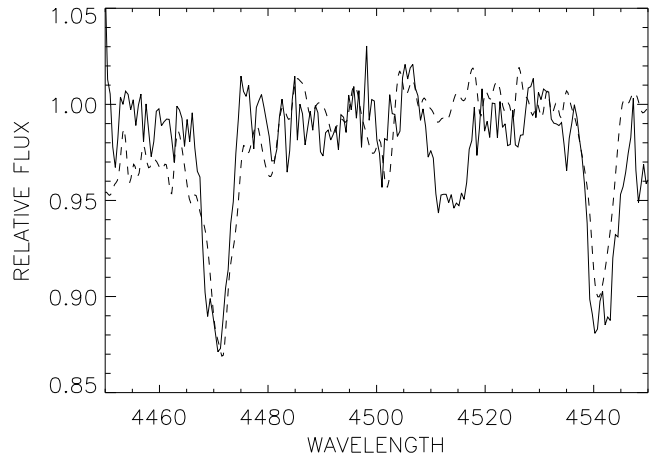}}
   \caption{The spectrum of A15 (O7\,Ib; solid line) compared to that of A11
     (O7.5\,Ib-II(f), dotted line). The strong \ion{N}{iii}~4511\,--\,14\AA\ feature
     and the high value of $\epsilon$ derived from the model fits for A15
     indicate advanced chemical evolution. The high rotational
     velocity and the underluminosity of this star all indicate
     anomalous evolution, perhaps due to mass transfer in a close binary. }
              \label{fig:comp}
    \end{figure}

\begin{figure}
   \centering
   \resizebox{\columnwidth}{!}{\includegraphics[angle =-90]{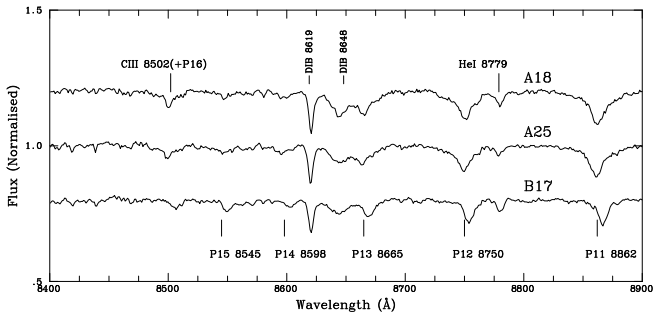}}
   \caption{$I$-band spectra of three targets observed with the WHT in
     August 2007. A25 does not show clear evidence for any Paschen
     line beyond Pa~13 and has weak \ion{C}{iii}~8502\AA. It is hence
     most likely an O8-9\,V star. A18 is clearly more luminous and
     probably earlier. Note the important radial velocity shifts in the
     lines of B17, this time in the opposite sense to those in the
     blue spectrum in Fig.~\ref{wht06}.}
              \label{fig:iband}
    \end{figure}

\begin{figure}
   \centering
   \resizebox{\columnwidth}{!}{\includegraphics[angle=-90]{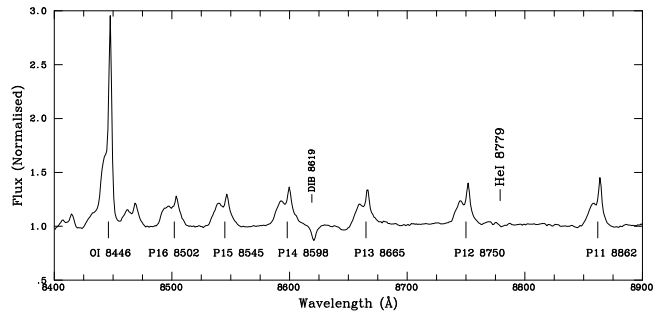}}
   \caption{$I$-band spectrum of B10. The data available identify it
     as an early ($\leq$ B2) Be star, but do not allow us to decide
     whether it is a classical Be star or a PMS Herbig Be object. The
     enormous strength of \ion{O}{i}~8446\AA, though, is unusual for a
     classical Be star and points to the second option.}
              \label{fig:be}
    \end{figure}

 Likewise, the He abundance by number relative to H plus He,
 $\epsilon$, is set initially to the standard value $\epsilon = 0.09$
 for  all stars, and varied to obtain better fits. Only A15 needed a
 higher value to fit the observed spectrum. The high value required by
 A15, $\epsilon= 0.25$, points to an evolved object, 
consistently with its low gravity and its strong N
spectrum. Fig.~\ref{fig:comp} shows a comparison of the  spectra of A11
and A15 (both O7 supergiants), around 4500\AA, where we can see the
strong \ion{N}{iii}~4511\,--\,14\AA\ feature in the spectrum of A15 . Note also that the projected
rotational velocity is remarkably high ($245\:{\rm km}\,{\rm s}^{-1}$)
for an evolved object (which is assumed to have 
lost significant amounts of angular momentum). The values derived
suggest that the 
evolution of this object has been anomalous (perhaps as a consequence of 
binary evolution) as, in addition, it appears underluminous and undermassive for
its spectral type.

 %______________________________________________ 
%
%______________________________________________ 
%

The parameters derived are given in
Table~\ref{params} and correspond very well to  the spectral types
derived in most cases. 
Errors in the stellar parameters are estimated at $\delta T_{{\rm eff}}=
\pm1500\:{\rm K}$,
$\delta \log g= \pm0.2$ and $\delta(\log \dot{M})= \pm0.3$ for the low
resolution observations and slightly lower in $T_{{\rm eff}}$ ($\pm1000$~K) 
and $\log g$ ($\pm0.15$) for the higher resolution data. Absolute
luminosities, radii and masses have been calculated from the $K_{{\rm
    S}}$ magnitude, assuming $DM=10.8$ ($d=1.4$~kpc), after \citet{hanson},
following the method discussed in
Section~\ref{sec:hr}. In the case of A38, the value $M_{V}=-3.7$ is
more than half a magnitude fainter than expected for the spectral
type, resulting in the low derived mass and luminosity. 

 Note that, in order to compare the absolute astrophysical parameters
 derived for these objects to those in previous works
 \citep{her99,her02}, they must be reduced to the same distance,
 as previous works assumed the canonical $DM=11.2$.

 %______________________________________________ 
%
%______________________________________________ 
%

Projected rotational velocities could not be determined for the stars
observed in 2006 due to the low 
resolution; the instrumental profile dominates the line broadening for 
the metals (H and He lines are broadened by the Stark
profile). However, this provides an upper limit, as instrumental
broadening dominates in all our objects, allowing us to ascertain that
they all rotate with $v \sin i<160\:{\rm km}\,{\rm s}^{-1}$.

 %______________________________________________ 
%
%______________________________________________ 
%

\section{Discussion}

\subsection{Completeness}
 
Our results confirm the enormous success of \citet{com02} at
identifying reddened OB stars. Only one of the candidates turns out to
be an interloper. The important point, however, is estimating whether
these objects are members of Cyg~OB2. The line of sight in this
direction runs parallel to the Local Arm, and populations at
different distances may lie projected together. While it is
extremely unlikely that early O-type might be found far away from
massive clusters or associations, except for a few runways
\citep[cf.][]{dewit05, vdb04}, less massive stars will certainly be found
if one looks through a Galactic Arm. 

%______________________________________________ 
%
   \begin{figure}
   \centering
   \resizebox{\columnwidth}{!}{\includegraphics{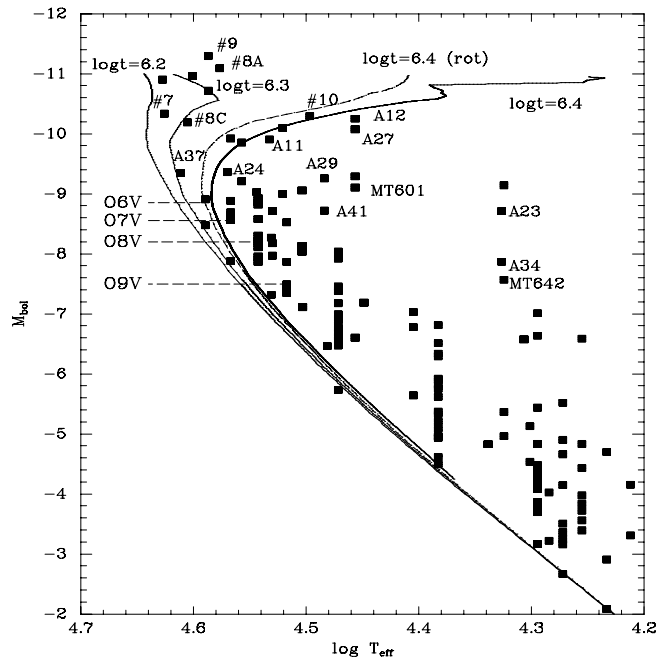}}  
   \caption{Semi-observational HR diagram for Cyg~OB2, based on published
     spectral types and 2MASS $JHK_{{\rm S}}$ photometry (see text for
   details). Continuous lines are non-rotating isochrones for $\log
   t=6.2$, $6.3$ and $6.4$ from \citet{schaller}. The dashed line is the
   $\log t=6.4$ isochrone in the  high-rotation models
   \citep{mm03}.}
              \label{fig:hr}
    \end{figure}
%______________________________________________ 
%

In this sense, our sample appears rather different from that
of \citet{hanson}, who observed only candidates which were bright in
$B$, most of which turned out to be foreground B-type stars. Even
though this is not surprising for the August 2007 sample, which was
selected on the basis of bright $K$ magnitudes, it is more striking
for the first (Loiano) sample, which simply consists of objects somewhat
fainter in $B$ than those observed by \citet{hanson}. This suggests
that, even though extinction is clearly variable across the face of
Cyg~OB2, on average, there is a range of extinctions where we can find
members, and this translates into a range of magnitudes (according to
spectral type). This range has been estimated as $4\la A_{V}\la 7$ by
previous authors \citep[e.g.,][]{mt91}, and may extend to somewhat
higher values when stars from the list of \citet{com02} are added.

Because of this, we suggest  that there  cannot be many more O-type members
amongst the candidates given by \citet{com02}, though there must be
some (for example, perhaps A17, see below). Very few of them are likely to be
intrinsically bright (and
so  very massive, evolved) members of the cluster. Based on their
$(J-K_{{\rm S}})$ colours and $K_{{\rm S}}$ magnitudes, only A4 and A8 might be
sufficiently bright (intrinsically) to be obscured O-type giants or
supergiants. 

In addition, our data reveal three new evolved O-type stars (A11, A15 and A24),
which help define the main-sequence turn-off of the association. A11
is of particular interest, as it lies very close to the Blue
Hypergiant (BHG) candidate \#12, in
what likely is the most obscured part of the association. We classify
it O7.5\,Ib-II(f), as it almost looks evolved enough to be a
supergiant, and its analysis indeed shows that it is a very massive
star. \cite{com02} suggest it may be the counterpart to the
X-ray source 1E~2023043+4103.9. Another candidate from \citet{com02}, 
A17, lies very close 
to it. It has very similar IR colours, but is two magnitudes fainter
in $K$. This is most likely a late-O/early-B main-sequence member.

\subsection{Clustering}

\begin{figure*}
   \centering
   \resizebox{\textwidth}{!}{\includegraphics{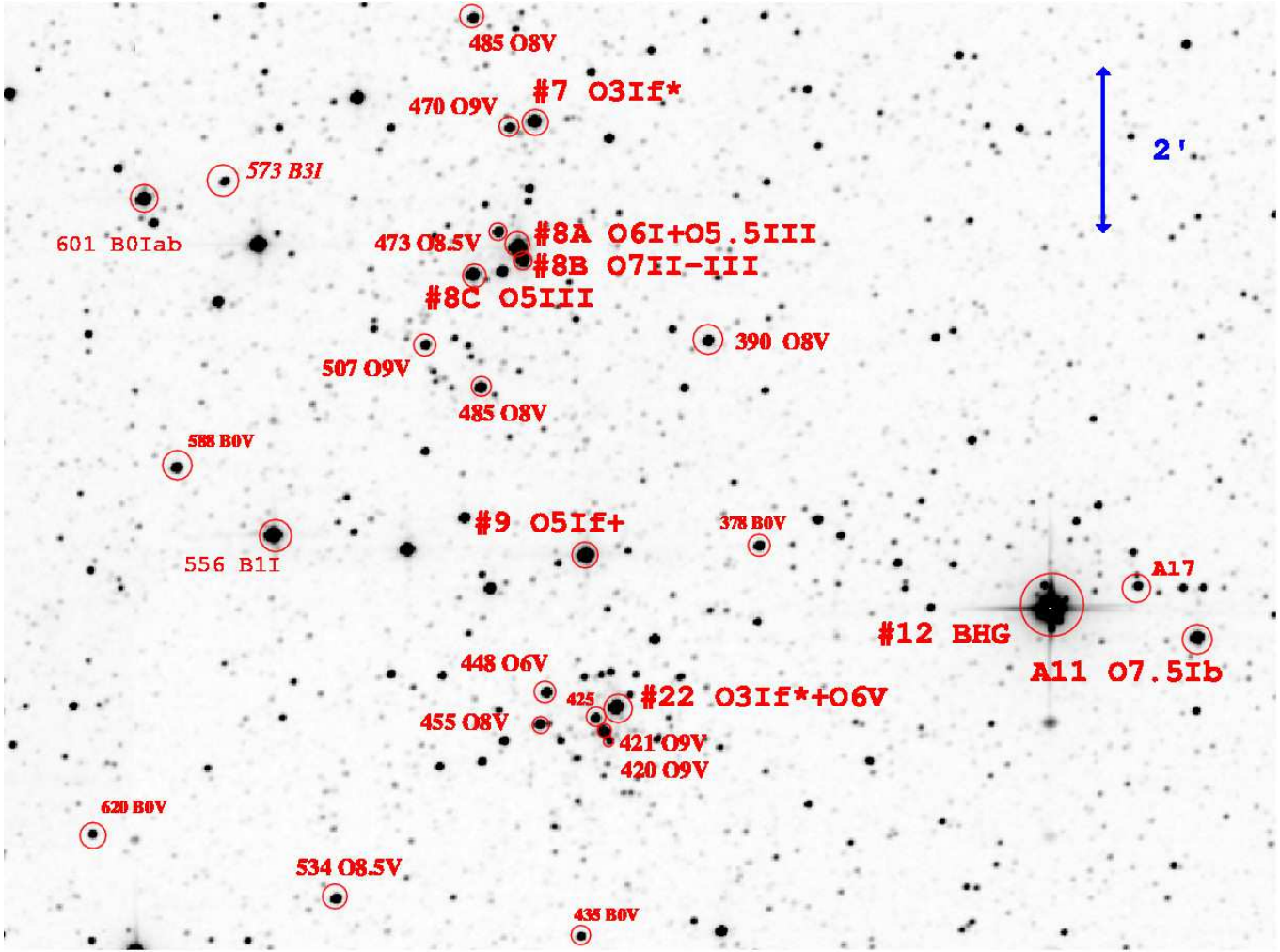}}
   \caption{A 2MASS $K_{{\rm S}}$ image of the central region of
     Cyg~OB2. All catalogued stars more massive than B0\,V are
     marked. MT573 (B3\,I) is obviously a background object, while the
   connection of MT556 (B1\,I) and MT601 (B0\,Iab) to the association
   is unclear. They 
   may represent an older ($\sim8$~Myr) population at about the same
   distance. The two cluster-like groupings identified by \citet{bica03}
   stand out in the image. Object~1 is centred on MT425 and contains
   \#22, while Object~2 is centred on the three components of \#8.}
              \label{fig:core}
    \end{figure*}

Fig~\ref{fig:core} shows a 2MASS
$K_{{\rm S}}$ image of the central region of Cyg~OB2, containing the
two cluster-like groupings identified by \citet{bica03} and the area
around \#12. The two clusters are prominent
against the background. The field shown in Fig.~\ref{fig:core} is
$\sim12\arcmin\times12\arcmin$, corresponding to $\sim5$~pc at
1.4~kpc. The separation between Cluster~1 and Cluster~2
($\la6\arcmin$) is equivalent to ~2.5~pc, and so smaller than the
radius of relatively massive clusters in the Perseus Arm, such as
h~Per or NGC~663. 

%______________________________________________ 
%
%______________________________________________ 
%

The nine brightest stars in Cluster 2 have $(J-K)_{{\rm
    S}}=0.63\pm0.03$ (standard deviation) and eight of them have $(J-K)_{{\rm
    S}}$ between 0.59 and 0.63. This uniformity in reddening represents strong confirmation of
their association, also clear in the colour-magnitude diagram shown by
\citet{bica03}. Cluster~1, on the other hand, does not seem to have 
a uniform reddening, but we do not think that this is a strong
argument against its reality, in this region of
patchy obscuration.    

In addition to these two groups, and outside the field covered by
Fig.~\ref{fig:core}, there is another obvious region of stellar
overdensity - present in the data of 
\citet{kiminki} - surrounding star \#4 (O7\,III). It comprises MT213
(B0\,V), MT215 
(B2\,V), MT216 (B1.5\,V)  and MT221 (B2\,V). MT187 (B1\,V), MT227
(O9\,V), MT241
(B2\,V) and MT258 (O8\,V)  lie within  $3\arcmin$. MT187 and
MT221 are significantly ($>0.1$~mag) more reddened than the others, but the
other seven have $(J-K)_{{\rm S}}=0.49\pm0.03$, again strongly
  hinting at a real
physical association.

The presence of all these small groups over a large area suggests that
star formation has proceeded in small bouts in this region, perhaps
over an extended period of time. In spite of this, the bulk of the
population occupies positions in the HR diagram incompatible with a
very long period of star formation. Subclustering is seen in the
largest Galactic star forming regions, such as W49A \citep{ha05} or
W51 \citep{nan04}. Study of large stellar complexes in M51
\citep{bas05} shows that the age spread within different clusters is
$\la10$~Myr.    

\subsection{HR diagram and ages}
\label{sec:hr}

As discussed by \citet{hanson}, the main sequence in Cyg~OB2 extends
clearly down to O6\,V.  
Star \#22, classified O4\,III(f) by \citet{mt91} has been shown to be
a close double containing an O3\,If* supergiant and an O6\,V star
\citep{wal02}. 
MT516, classified as O5.5\,V((f)) by \citet{mt91}, was found to
have a rather low gravity by \citet{her99}. Indeed, the fact that
\ion{He}{ii}~4542\AA\ is somewhat stronger than \ion{He}{ii}~4686\AA\ shows
that this object is rather far away from the ZAMS.
% The same is true of A37, which \citet{hanson} classifies as
% O5\,V((f)).
Therefore this
star is likely better classified as O5.5\,III, joining \#8C and the
faint component of  \#8A as an
object still on the main 
sequence, but already showing some signs of evolution. Therefore the
age of the association would seem to be set by the fact that stars
more massive than O6\,V are already somewhat evolved, while O6\,V
stars are not. 

However, within a classical theory of stellar evolution, it is
difficult to see how this fits with the presence of O3 supergiants. In
order to address this question and also exploit the potential of
Cyg~OB2 as a laboratory, we have constructed an HR diagram utilising the
wealth of new spectral type determinations in this region. We have
used the 2MASS $JHK_{{\rm S}}$ magnitudes for all objects and their
spectral types (from \citealt{kiminki}, \citealt{hanson} or this work) in order
to place them in a semi-observational HR diagram. We have followed the
procedure used by, for instance, \citet{mas95}, but taking infrared
rather than optical magnitudes. We have resorted to $JHK_{{\rm S}}$
magnitudes partly because many stars of interest lack good $U$-band
photometry, but also because this allows a test of the usefulness of
infrared data to study obscured massive clusters.

From the spectral types derived, we have taken a $T_{{\rm eff}}$ and
bolometric correction $BC$,
using the calibration of \citet{mar05} for O-type stars and that
of \citet{hme84} for B-type stars (the two calibrations agree quite
well around B0; however, the possible existence of an artificial jump
between B0 and B1 has been noted by previous authors). We also take intrinsic
$(V-K)_{0}$ and $(J-K)_{0}$
colours from the calibration of \citet{wegner}. With the observed
$(J-K_{{\rm S}})$, we derive $E(J-K_{{\rm S}})$. As the reddening to
the association is 
known to be very close to standard \citep{hanson}, we simply calculate
$A_{K_{{\rm S}}}=0.67E(J-K_{{\rm S}})$. 

We then calculate $K_{0}=K_{{\rm S}}-A_{K_{{\rm S}}}-DM$, using
$DM=10.8$ from \citet{hanson}, and by adding $(V-K)_{0}$ and the $BC$,
arrive at a semi-observational $M_{{\rm Bol}}$. Fig.~\ref{fig:hr}
plots $M_{{\rm Bol}}$ against the $T_{{\rm eff}}$ derived from the
spectral classification. Superposed on it, are Geneva isochrones without
rotation for $\log t=6.2$ (1.5~Myr), $\log t=6.3$  (2~Myr) and $\log
t=6.4$ (2.5~Myr), as well as the rotating isochrone for $\log
t=6.4$. We tried to fit the 
data using the higher distance modulus ($DM=11.3$) obtained by
averaging spectroscopic distances
\citep{kiminki}, but this left all the stars well above the ZAMS. The
data used for this diagram are listed in Table~3.

%The
%spectroscopic DM is likely affected by the unrecognised presence of a
%substantial fraction of binaries with two similar components.

The diagram shows several notable features. There is a very well traced
main sequence extending to the O6\,V stars. As noted by
\citet{hanson}, A37 (O5\,V) is earlier than any other MS stars, but
seems to fit the sequence well. Around $\log T_{{\rm
    eff}}\sim4.3$, there lie a number of B stars well above the main
sequence, unlikely to be connected with the rest of the
population. Objects like MT642 (B1\,III), A23 (B0.7\,Ib) or A34
(B0.7\,Ib) are very probably not members of Cyg~OB2.  

There are two evolutionary sequences that seem to turn off the main
sequence. The one below the isochrones is formed by MT138 (O8\,I),
A32 (O9.5\,IV), A41 (O9.7\,II) , A29 (O9.7\,Iab), A36 (B0\,Ib), and
MT601 (B0\,Iab). Though this sequence may provide a decent fit to the $\log
t=6.7$ (5~Myr) isochrone, the random distribution in luminosity class
suggests that this is not a real evolutionary sequence, but simply the
projection 
of a number of luminous stars situated at slightly different
distances. It is worthwhile mentioning, though, that many of these
objects have  distance moduli comparable to that of the main Cyg OB2
association. 

In contrast, the sizable population of evolved stars lying around the $\log
t=6.4$ isochrone seems to form a much more coherent group. Moving
along the isochrone, we have A24 (O6.5\,III) 
and \#4 (O7\,III), \#8B (O7\,II-III), A11 (O7.5\,Ib-II), A20 (O8\,II),
\#10 (=MT632, O9.5\,Ia), A12 (B0\,Ia) and A27 (B0\,Ia). The excellent
progression 
in luminosity class with spectral type  strongly supports the
hypothesis that these objects 
are really following the isochrone. Only two stars with accurate
spectral types do not fit this evolutionary sequence: one  is MT771
(O7\,V), which appears as bright as \#4. This is easily explained by the
fact that it is a double-lined spectroscopic binary with two similar
components \citep{kiminki}. The other one is A15 (O7\,Ib),
which is $\sim0.6$~mag fainter than expected. As mentioned, the
analysis of its spectrum reveals very high He and N abundances, and a very
low mass for its spectral type. Therefore this is indeed a peculiar
object, perhaps the product of mass transfer in a close binary, as
\#5 and B17 are also likely to be.

The distribution of main sequence stars, the main-sequence turn-off and
the sequence of  evolved stars strongly supports an age $\sim 2.5$~Myr
for the bulk of Cyg~OB2. According to the calibration of
\citet{mar05}, this implies that stars up to $\sim35\,M_{\sun}$ are
still close to the ZAMS, while more massive stars are already more
evolved. However, it is obvious that the brightest
stars in the association fall well above the adopted isochrone.

Given the extent of Cyg~OB2, the possibility of a spread in ages
cannot be excluded and may even seem logical. Indeed, in a recent
paper, \citet{drew08} have shown an important concentration of A-type
stars to the South of the bulk of the O-type stars. In order to be at
the same distance as the O-star association, these A-type stars must
be part of a 5\,--\,7~Myr population.

Does the presence of O3 supergiants indicate the existence of an even
younger population? Certainly this possibility cannot be excluded, but
it is worth taking in consideration two points:

\begin{itemize}

\item If there is an age difference, we would expect to find some
sort of spatial segregation between the older and younger
population, but  this is not evident in the data.  The area shown in
Fig.~\ref{fig:core} contains most of the
earliest objects, but also A11 (O7.5\,Ib-II) and \#12. The moderately
evolved \#8B (O7\,II-III) falls just in the middle of Cluster 2, which
contains three of the early objects. 

\item Stars more massive than the O6\,V objects still in the main
  sequence appear as either earlier-type supergiants (Of$^{*}$ stars)
  or intermediate luminosity O6--7 stars. In other words, if there is
  a younger population, all its members appear as Of$^{*}$ stars just
  now. 
\end{itemize}

 At
the estimated age, it is unlikely that any stars might have undergone
supernova, and indeed no supernova remnant is seen in the area
\citep{pas02}.  All the stars in Fig.~\ref{fig:hr} occupy positions in the
HR diagram compatible 
with being still in the hydrogen core-burning phase. The BHG candidate
\#12 may be past this phase and there are 5 Wolf-Rayet stars in the
region that have been proposed as possible members \citep{pas02}. The
WC stars WR~144 and WR~146 may actually be the descendants of the most
massive stars in the association \citep{pas02}. The fact that some of
the most massive stars appear as O3\,If supergiants, 
while others are moving towards the red part of the HR diagram (or
seem already to be locked in an LBV phase, like \# 12) is
highly suggestive of the idea that not all very massive stars evolve
in the same way. 

There is, however, ample evidence suggesting that star formation has
been going on for quite some time in a large area around the
recognizable core of Cygnus OB2. Indeed many of the evolved massive
stars that are unlikely to belong to the current generation of massive
stars lie at
approximately the same distance and could belong to an older ($\sim7$~Myr)
generation,  associated with the young A-type stars
detected by \citet{drew08}. In the massive association 30 Dor,
\citet{wb97} also find a population of OB stars somewhat older
(4\,--\,6~Myr) than those in the main cluster R136 scattered
across the entire complex. Likewise, \citet{mok07} find ages of $7.0\pm1.0$
and $3.0\pm1.0$~Myr for the associations LH9 and LH10, in the giant
\ion{H}{ii} region N11 in the LMC. Their data are
consistent with LH9 having triggered star formation in LH10.

\section{Conclusions}

 Though the candidate sample of \citet{com02} contains a
       high fraction of likely non-members, as discussed by
       \citet{hanson}, it has also allowed the detection of a number
       of obscured O stars and very luminous B0\,Ia supergiants very
       likely to be members of Cyg~OB2.

 When these objects are included in the HR diagram, it
       becomes clear that there is a sequence of moderately evolved
       stars detaching from the main sequence exactly at the position
       where we stop seeing luminosity class V objects, i.e., around
       O6\,V. These two facts combined support an age of
       $\sim2.5$~Myr for the bulk of the association.

 The HR diagram presented in Figure~\ref{fig:hr} contains the largest
 number of Cyg~OB2 members ever displayed in such a diagram. It contains
 $\sim 50$ stars that may have started their lives as main-sequence
 O-type stars and only  a few of these are unlikely to be members. Unless a
 population of extremely obscured O-type stars is lying at fainter
 magnitudes than probed by 2MASS, the total number of O-type stars in
 the association is
 almost certain to be in the 60\,--\,70 range. 

 The number of stars that have already left the main
 sequence and lie above the O6\,V members that define the
 turn-off is more securely determined.  If the main association is basically co-eval, these
 represent the subset of stars  that were originally more massive than
 $35\,M_{\sun}$. Counting \#12, which is not shown 
 in Fig.~\ref{fig:hr} because of its claimed spectral variability
 \citep{kiminki}, there are 21 such stars. The evolved
 interacting binaries \#5 and B17 (not in Fig.~\ref{fig:hr}) should be
 counted too (perhaps doubly). The resulting number is certainly
 only a lower limit. Apart from possible unrecognised close doubles and
 binaries, at least two of the  Wolf-Rayet stars in the area are likely to be
 descendants of very massive stars \citep{pas02}. Also, \cite{com07}
 identify BD~$+53\degr$3654 as a likely runaway O4\,If member of the
 association. Therefore, we have identified a population of at least 25
 stars that were originally more massive than $35\,M_{\sun}$. Given the
 uncertainties - in particular the very
 high binary fraction \citep{kiminki} - we refrain from trying to
 derive a total mass for the association by assuming an IMF.

 The brightest members, with spectral types in the O3--O5
        range, may technically be considered blue stragglers. Though a
       real age difference cannot be ruled out, it does not seem to be
        borne out by the spatial 
        distribution of stars, perhaps suggesting that we are seeing stars of
        similar mass evolve in very different ways.

Three luminous supergiants (\#10 O9.5\,Ia, A12 B0\,Ia and
         A27 B0\,Ia) seem to follow the 2.5~Myr isochrone and so appear to be
         the descendants of stars more massive than
         $\sim40\,M_{\sun}$. This is in agreement with an initial mass
         estimate of $48\,M_{\sun}$ for \#10 \citep{her02}, which may
         have to be slightly reduced if the lower $DM=10.8$ is
         adopted. These objects will probably soon reach the LBV
         instability, which \#12 is perhaps already encountering. A large
         population of 
         O9--B1\,Ia supergiants descended from stars with
         $M_{*}\approx35\,M_{\sun}$ is found in the older
         ($\sim4.5$~Myr) cluster Westerlund~1 together with a number
         of LBVs and Yellow Hypergiants \citep{clark05}. 

In summary, even if Cyg~OB2 falls short of the proposed 100 O-type
stars by a factor of $\sim 2$, its nuclear region still represents one
of the most massive collections of early-type stars known in the
Galaxy and its relatively low reddening cements its value as a
laboratory for the study of their properties. 

\begin{acknowledgements}

We thank Vanessa Stroud for help with the 2007 run and reduction of
some spectra.
      
During most of this work, IN was a researcher of the programme {\em
  Ram\'on y Cajal}, funded by the Spanish Ministerio de Educaci\'on y
Ciencia and the University of Alicante, with partial 
support from the Generalitat Valenciana and the European Regional
Development Fund (ERDF/FEDER).
This research is partially supported by the MEC under
grants AYA2005-00095, AYA2004-08271-C02-01, AYA2007-67456-C02-01 and
CSD2006-70 and by the Generalitat Valenciana
under grant GV04B/729.   

The G.D. Cassini telescope
is operated at the Loiano Observatory by the Osservatorio Astronomico di
Bologna.
 The WHT is operated on the island of La
Palma by the Isaac Newton Group in the Spanish Observatorio del Roque
de Los Muchachos of the Instituto de Astrof\'{\i}sica de Canarias. The
June 2006 observations were taken as part of the service programme
(programme SW2005A20).

This research has made use of the Simbad data base, operated at CDS,
Strasbourg (France). This publication makes use of data products from
the Two Micron All 
Sky Survey, which is a joint project of the University of
Massachusetts and the Infrared Processing and Analysis
Center/California Institute of Technology, funded by the National
Aeronautics and Space Administration and the National Science
Foundation.

\end{acknowledgements}

\end{document}